\documentclass[prb,twocolumn,superscriptaddress]{revtex4}
\usepackage{mathrsfs,amsmath,amssymb,paralist,epsfig}
\usepackage{psfrag}
\usepackage{graphicx,bm}

\def\bea{\begin{eqnarray}}
\def\eea{\end{eqnarray}}
\def\pp{\parallel}
    \def\<{\langle}         \def\>{\rangle}

\def\sgn{\,\mathrm{sgn}}

  \def\V0{{\mathbf 0}}

  \def\B0{{\mathbf 0}}

\def\be{\begin{equation}}       \def\ee{\end{equation}}
\def\bea{\begin{eqnarray}}      \def\eea{\end{eqnarray}}
\def\nn{\nonumber}

\begin{document} 

\title{$F$-wave pairing of cold atoms in optical lattices}

\author{Wei-Cheng Lee}
\email{leewc@physics.ucsd.edu}
\affiliation{Department of Physics, University of California, San Diego,
CA 92093}
\author{Congjun Wu}
\email{wucj@physics.ucsd.edu}
\affiliation{Department of Physics, University of California, San Diego, 
CA 92093}
\author{S. Das Sarma}
\email{dassarma@umd.edu}
\affiliation{Condensed Matter Theory Center, Department of Physics, 
University of Maryland, College Park, MD 20742}

\begin{abstract} 
The tremendous development of cold atom physics has opened up a whole
new opportunity to study novel states of matter which are not easily
accessible in solid state systems.
Here we propose to realize the $f$-wave pairing superfluidity of 
spinless fermions in the $p_{x,y}$-orbital bands of the two dimensional
honeycomb optical lattices. 
The non-trivial orbital band structure rather than strong correlation
effects gives rise to the unconventional pairing with the nodal lines of 
the $f$-wave symmetry.
With a confining harmonic trap, zero energy Andreev bound states
appear around the circular boundary with a six-fold symmetry. 
The experimental realization and detection of this novel pairing state 
are feasible.
\end{abstract} 

\maketitle

%*******************************************************************
\section{Introduction}
The study of unconventional Cooper pairing states has been a major 
subject in condensed matter physics for decades.
In addition to the isotropic $s$-wave pairing, many unconventional 
pairing states have been identified \cite{sigrist1991}.
For example, different types of $p$-wave pairing states were found 
in the superfluid $^3$He systems \cite{leggett1975,osheroff1997}
including the anisotropic chiral $A$-phase and the isotropic $B$-phase.
Evidence of the $p$-wave paring was also found in the
ruthenate compound of Sr$_2$RuO$_4$ \cite{nelson2004,kidwingira2006}.
The $d$-wave pairing states are most convincingly proved in the high-$T_c$
cuprates with  phase sensitive measurements including
Josephson tunneling junction \cite{harlingen1995,tsuei2000} 
and zero energy Andreev bound states \cite{hu1994,deutscher2005}.
Many heavy fermion compounds exhibit evidence of unconventional Cooper
pairing such as UPt$_3$,  UBe$_{13}$, and CeCoIn$_5$ with nodal points
or lines \cite{joynt2002,maclaughlin1984,izawa2001,strand2009}.
However, conclusive experimental evidence to determine
pairing symmetries is still lacking.
These unconventional pairing states are driven by strong 
correlation effects, which brings significant difficulties 
for theoretical analysis and prediction.
It would be great to find novel unconventional pairing
mechanisms easier to handle.

On the other hand, cold atom physics has recently become an 
emerging frontier for condensed matter physics 
\cite{dalfovo1999,leggett2004,bloch2008,giorgini2008}.
In particular, the $s$-wave pairing superfluidity of fermions 
through the Feshbach resonances has become a major research focus.
The Bose-Einstein condensation (BEC) and Bardeen-Cooper-Schrieffer 
(BCS) crossover has been extensively investigated 
\cite{chin2008,chin2004,regal2004,zwierlein2004,partridge2005}.
Naturally, searching for unconventional Cooper pairing states in cold
atom systems is expected to stimulate more exciting  physics.
For example, the $p$-wave pairing states have been proposed
by using the $p$-wave Feshbach resonances \cite{ho2005,gurarie2007,cheng2005},
which, however, suffer from a drawback of heavy particle loss.
The unconventional Cooper pairing states with cold atoms have 
not been realized yet \cite{zhang2004,gaebler2007,fuchs2008,inada2008}.

In this article, we propose a novel $f$-wave pairing state of
spinless fermions in the cold atom optical lattices.
This unconventional pairing arises from the non-trivial band 
structure of the $p_{x,y}$-orbital bands in the honeycomb
optical lattices combined with a conventional attractive interaction.
The internal orbital configurations of the Bloch wave band eigenstates vary
with the crystal momenta,  resulting in an $f$-wave angular
form factor for the pairing order parameters.
Along three high symmetry lines in the Brillouin 
zone whose directions differ by 120$^\circ$ from each other, the intra-band
pairing order parameters are exactly suppressed to zero.
The unconventional nature of this pairing exhibits in the
appearance of the zero-energy Andreev bound states at the circular 
boundary with imposing a confining trap.
Since no strong correlation effects are involved, our analysis below
is well-controllable.
This is a novel state of matter, which to our knowledge has not 
been unambiguously identified before neither in condensed matter nor
in cold atom systems, thus this result will greatly enrich the study 
of unconventional pairing states.

This article is organized as follows.
In Section \ref{sect:band}, we explain the orbital structure
of the Bloch wave eigenstate in detail.
In Section \ref{sect:fwave}, we show that the $f$-wave Cooper
pairing naturally arises from a conventional on-site Hubbard attraction.
In Section \ref{sect:andreev}, we discuss the zero energy
Andreev bound state at the edge of the confining trap,
which provides a phase sensitive evidence.
In Section \ref{sect:detect}, we discuss the experiment
realization and detection.
Conclusions are given in Section \ref{sect:conclusion}.

%*********************************************************************
\section{The orbital structure of the $p$-band Bloch wave functions}
\label{sect:band}

The honeycomb optical lattice was experimentally constructed quite some
years ago by using three coplanar blue detuned laser beams whose wavevectors
$\vec k_i ~(i=1,2,3)$ differ by 120$^\circ$ from each other
\cite{grynberg1993}.
After the lowest $s$-orbital band is fulfilled, the next active
ones are $p_{x,y}$-orbital bands lying in the hexagonal plane. 
Different from the situation in graphene whose $p_{x,y}$-orbital bands
strongly hybridize with the $s$-orbital bands and are pushed away from the 
Fermi surface, the hybridization between the $p_{x,y}$ and $s$-orbitals 
in the optical honeycomb lattice is negligible.
The $p_z$-orbital band can be pushed to higher energies and thus unoccupied
by imposing strong confinement along the $z$-axis.
The $p_{x,y}$-orbital bands exhibit the characteristic features of
orbital physics, {\it i.e.}, spatial anisotropy and orbital degeneracy.
This provides a new perspective of the honeycomb lattice, 
and is complementary to the research focusing on the
$p_z$-band system of graphene which is not orbitally active.
The novel physics which does not appear in graphene includes the flat 
band structure\cite{wu2007}, the consequential non-perturbative 
strong correlation effects (e.g. Wigner crystal \cite{wu2007a}
 and ferromagnetism\cite{zhang2008}), 
frustrations in orbital exchange\cite{wu2008}, and the 
quantum anomalous Hall effect\cite{wu2008a}.

We employ the $p_{x,y}$-orbital band Hamiltonian studied in Ref.
\cite{wu2007,wu2007a,wu2008,wu2008a} as
\bea
H_0=t_\parallel\sum_{\vec{r}\in A, i=1,2,3}\{ p^\dagger_{\vec{r},i} p_{\vec{r}+a\vec{e}_i,i} 
+ {\rm h.c.}\}-\mu\sum_{\vec{r}\in A\oplus B} n_{\vec{r}},
\label{eq:ham0}
\eea
where $\hat{e}_{1,2,3}$ are unit vectors from one site in sublattice $A$ to 
its three neighboring sites in sublattice $B$ defined as 
$\hat{e}_{1,2}=\pm\frac{\sqrt{3}}{2}\hat{e}_x+\frac{1}{2}\hat{e}_y,
\hat{e}_3=-\hat{e}_y$;
$p_i\equiv (p_x\hat{e}_x+p_y\hat{e}_y)\cdot\vec{e}_i$
is the $p$-orbital projected onto the bond along the direction of
$\hat e_i$ and only two of them are linearly independent;
$n_{\vec r}=n_{\vec r,x}+n_{\vec r, y}$ is the particle number at site $\vec r$;
the $\sigma$-bonding $t_\parallel$ describes the hopping between
$p$-orbitals on neighboring sites parallel to the bond
direction and is rescaled to 1 below; $a$ is the nearest neighbor distance.
$t_\parallel$ is positive due to the odd parity of $p$-orbitals.
Eq. \ref{eq:ham0} neglects the $\pi$-bonding $t_\perp$ describing
the hopping between $p$-orbitals perpendicular to the bond direction.
Due to the high spatial anisotropy of the $p$-orbitals, $t_\perp/ t_\pp\ll 1$.
To confirm this, we have fitted $t_\perp$ and $t_\pp$
from a realistic band structure calculation for the
sinusoidal optical potential 
\bea
V(\vec r)=V_0 \sum_{1\le i<j \le 3} 
\cos \{(\vec k_i -\vec k_j)\cdot \vec r \}.
\eea
With a moderate potential depth of $V_0/E_R=12$, 
$t_\perp/t_\pp$ is already driven to $1\%$
with $t_\pp\approx 0.375 E_R$ 
where $E_R=\frac{\hbar^2 k^2}{2M}$ is the recoil energy
and $M$ is the atom mass.
Further increasing $V_0/E_R$ decreases $t_\perp/t_\pp$ even more.

Eq. \ref{eq:ham0} has a chiral symmetry, {\it i.e.}, under
the transformation of $p_{\vec r,x(y)}\rightarrow 
-p_{\vec r,x(y)}$ only for sites in sublattice A but not
for sites in sublattice B, $H_0\rightarrow -H_0$.
Thus its four bands have a symmetric spectra
respect to the zero energy as
\bea
E_{1,4}=\mp\frac{3}{2}t_\parallel,\,\,\,E_{2,3}=
\mp\frac{t_\parallel}{2}\sqrt{3+2\sum_{i<j} \cos\vec{k}\cdot (
\hat{e}_i-\hat e_j)}. \nn \\
\eea
The Brillouin zone (BZ) is a regular hexagon with the 
edge length $\frac{4\pi}{3\sqrt 3 a}$.
Two middle dispersive bands $2$ and $3$ have two non-equivalent Dirac
points $K (K^\prime)=(\pm\frac{4\pi}{3\sqrt 3 a},0)$ with a band 
width of $\frac{3}{2} t_\parallel$. 
The bottom and top bands are flat.
We define the four component annihilation operators in momentum space 
as $\hat \phi(\vec k)=[\hat p_{A,x}(\vec k),
\hat p_{A,y}(\vec k), \hat p_{B,x} (\vec k), \hat p_{B,y} (\vec k )]^T$.
In this basis, the eigen-operator $\hat \psi_m (\vec k)$ for 
band $m$ can be diagonalized as
$\hat \psi_m (\vec k) = \hat \phi_n (\vec k) U_{nm} (\vec k)$ where
$U(\vec k)$ is a $4\times 4$ unitary matrix.
The phase convention for band eigenvectors, {\it i.e.}, 
each column of $U (\vec k)$,
is  conveniently chosen as $R_{\frac{\pi}{3}} \hat \psi_m (\vec k) 
R^{-1}_{\frac{\pi}{3}} 
=\sgn(m) \hat \psi_m(\vec k^\prime)$ with $\sgn(m)=-$ for
$m=1,2$ and $\sgn(m)=+$ for $m=3,4$, where the
symmetry operation $R_{\frac{\pi}{3}}$ is the $60^\circ$ rotation 
around a center of the hexagonal plaquette
and $\vec k^\prime =R_{\frac{\pi}{3}} \vec k$.
The analytical form of $U_{nm}(\vec k)$ is given in Appendix
\ref{sect:appendixA}.

%----------------------------------------------------------------------
\begin{figure}
\centering\epsfig{file=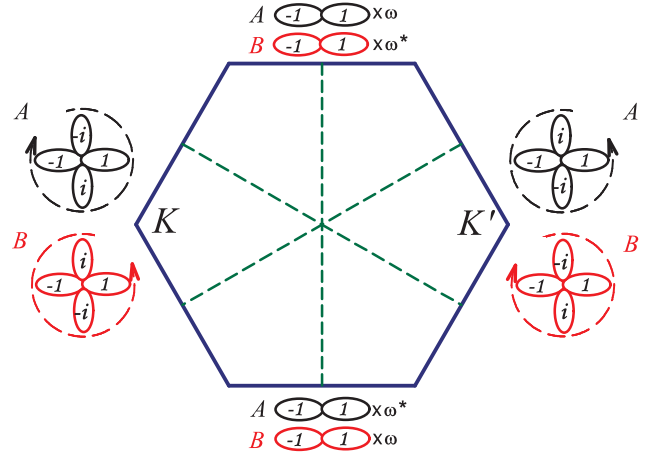,clip=1,width=\linewidth,angle=0}
\caption{(Color online)
Orbital configurations of the lower dispersive band 2 at high
symmetry points and lines in the BZ.
The time-reversal partners $\psi_2(\pm\vec k)$
only involve the same polar orbital if $\vec k$ is along the three
middle lines, and take the complex orthogonal orbitals of $p_x\pm ip_y$
with opposite chiralities at $\vec K$ and $\vec K^\prime$. 
$\omega=e^{i\frac{\pi}{6}}$. 
}
\label{fig:eigenvectors}
\end{figure}
%----------------------------------------------------------------------

The orbital configurations of the band eigenvectors have 
interesting patterns as depicted in Fig. \ref{fig:eigenvectors},
which arise from the lattice $D_{6h}$ symmetry.
Only the lowest dispersive band ($m=2$) is plotted as an example. 
The lowest flat band ($m=1$) has the same symmetry structure except 
having different orbital 
polarization directions.
The remaining two can be obtained by performing the operation of the chiral symmetry.
Six lines in the BZ possess  reflection symmetry, {\it i.e.}, 
three passing the middle points of the opposite 
edges and three passing the opposite vertices 
of $\vec K$ and $\vec K^\prime$.
The eigenvectors of each band along these lines should be either 
even or odd with respect to the corresponding reflection.
For example, for the reflection respect to the $y$-axis,
the orbitals transform as 
\bea
\hat p_{A(B),x}\rightarrow -\hat p_{A(B),x}, \ \ \
\hat p_{A(B),y}\rightarrow \hat p_{A(B),y},
\eea
thus 
the eigen-operators $\hat \psi_m (\vec k)$ with $\vec k \parallel \hat y$ 
must be either purely $\hat p_y$ (even)  or $\hat p_x$ (odd). 
For the other two middle lines, the corresponding eigenvectors
are obtained by performing the $\pm 120^\circ$ rotation.
For the time-reversal partners $\hat \psi_m(\vec k)$ and 
$\hat \psi_m(-\vec k)$ along these lines, they only take the same
real polar orbitals.
On the other hand, the reflection respect to the $x$-axis
gives rise to the transformation
\bea
\hat p_{A(B),x} \rightarrow \hat p_{B(A),x}, \ \ \,
\hat p_{A(B),y}\rightarrow -\hat p_{B(A),y}.
\eea
Furthermore, $\vec K$ and $\vec K^\prime$ have three-fold rotational symmetry.
Combining two facts together, $\hat \psi_m(\vec K)$ 
should be of $p_x + ip_y$ for sites in one of the sublattices
and $p_x-i p_y$ for sites in the other sublattice.
Its time-reversal partner $\hat \psi_m (\vec K^\prime)$ has the opposite
chiralities in both sublattices respect to those at $\vec K$.
In other words, 
the orbital configurations of $\hat \psi_m (\vec k)$
are linear-polarized at the middle points of the BZ edge,
changes to circularly-polarized at the vertices, 
and elliptically polarized in between.

%%%%%%%%%%%%%%%%%%%%%%%%%%%%%%%%%%%%%%%%%%%%%%%%%%%%%%%%%%%%%%
\section{F-wave Cooper pairing}
\label{sect:fwave}

Next we introduce the on-site attractive interaction term
between spinless fermions as
\be
H_{int}=-U\sum_{\vec{r}} n_{\vec{r},x}n_{\vec{r},y},
\label{eq:int}
\ee
where $U$ is positive.
We perform the mean-field decomposition for $H_{int}$ in the pairing channel as
\bea
H^{pairing}_{int}&=&- \sum_{\vec r_{A,B}}\big \{\Delta^*_A p_{\vec r_A, y}
p_{\vec r_A, x} +\Delta^*_B p_{\vec r_B, y}
p_{\vec r_B, x} 
+ h.c \big\}, \nn \\
&=&\sum^\prime_{\vec{k}}\sum_{m,n=1}^4\,
\Delta^*_{nm}(\vec{k})\psi_n(\vec{k})\psi_m(-\vec{k})+h.c,
\label{eq:hmf1}
\eea
where the pairing order parameters in the $A$ and $B$-sublattices
are self-consistently defined as $\Delta^*_{A(B)}=U\langle
G|p^\dagger_{r_{A (B),x}}p^\dagger_{r_{A(B),y}}|G\rangle$;
$\langle G| ..|G\rangle$ means the average over the pairing
ground state; 
the summation of $\vec k$ only covers half of the BZ;
the multi-band pairing order parameters in momentum space
has a $4\times 4$ matrix structure as
\bea
\Delta_{nm}(\vec k)=U\langle G|\psi_n(\vec k) 
\psi_m (-\vec k) |G\rangle .
\eea
Under the rotation of $R_{\frac{\pi}{3}}$, it transforms as
$R_{\frac{\pi}{3}} \Delta_{nm} (\vec k) R_{\frac{\pi}{3}}^{-1} 
=\sgn(n)\sgn(m) \Delta_{nm} (\vec k^\prime)$.

In addition, we also introduce the mean-field decomposition for $H_{int}$ 
in the charge-density-wave (CDW) channel as:
\bea
H^{CDW}_{int} = \sum_{\vec r_{A,B},\tau=x,y} &(-U\,n-\frac{N}{2})
&p^\dagger_{\vec r_A, \tau} p_{\vec r_A, \tau}\nn\\
+ &(-U\,n+\frac{N}{2})& p^\dagger_{\vec r_B, \tau} p_{\vec r_B, \tau},
\eea
where 
\bea
n=\langle G|(n_{\vec{r}_A,x}+n_{\vec{r}_A,y}+n_{\vec{r}_B,x}+n_{\vec{r}_B,y})/2|G\rangle
\eea
is the total particle filling, and 
\bea
N=U\langle G|(n_{\vec{r}_A,x}+n_{\vec{r}_A,y}-
n_{\vec{r}_B,x}-n_{\vec{r}_B,y})|G\rangle
\eea
is the CDW order parameter.
Generally speaking, the CDW becomes important in large $U$ or close 
to half-filling ($n=1$) as will be discussed below.

%-------------------------------------------------------------------
\begin{figure}
\centering\epsfig{file=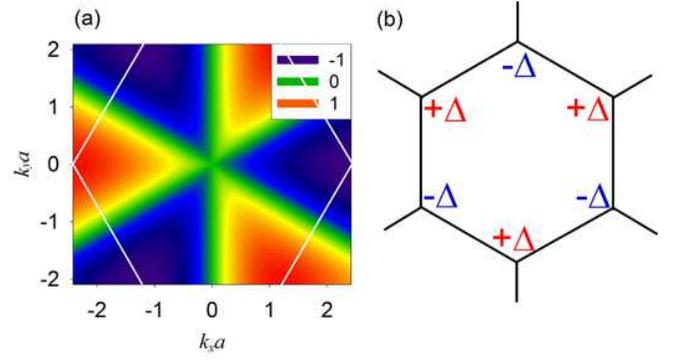,clip=1,width=\linewidth,angle=0}
\caption{
(Color online)
A) The $f$-wave pairing form factor $F(\vec k)$ for the 
intra-band pairing in momentum space. B) The $f$-wave pairing pattern in 
real space with $\Delta_A=-\Delta_B=\Delta$.
}
\label{fig:fwave}
\end{figure}
%--------------------------------------------------------------------

The intra-band gap functions $\Delta_{nn}$ can be calculated analytically 
as $\Delta_{nn}(\vec{k})= i(-)^n \frac{1}{2}(\Delta_A-\Delta_B) F(\vec{k})$
with the $f$-wave form factor of
\bea
F(\vec{k})=\frac{16}{\sqrt{3} N_0(\vec{k})}\sin\frac{\sqrt{3}}{2}
k_x\left[\cos\frac{\sqrt{3}}{2}k_x - \cos\frac{3}{2}k_y\right],
\label{eq:fwave}
\eea
where $N_0$ satisfies
\bea
N_0(\vec k)=\frac{8}{3}\big\{3-\sum_{1\le i<j\le 3} \cos \vec k \cdot 
(\hat e_i-\hat e_j)\big\}.
\eea
$\Delta_A$ can be fixed positive, and $\Delta_B=|\Delta_B| 
e^{i\Delta\theta}$ with a relative phase $\Delta \theta$.
The optimal $\Delta\theta$ takes the value of $\pi$, {\it i.e.},
$\Delta_A=-\Delta_B$, to maximize the intra-band pairings.
We have confirmed this in the explicit self-consistent mean-field
solution for Eq. \ref{eq:ham0} and Eq. \ref{eq:hmf1}.
Furthermore, the non-vanishing $\pi$-bonding $t_\perp$ term
can further stabilize this solution as a result of the odd parity 
of $\pi$-orbitals.

Now we discuss the pairing symmetry of this system. 
Since the fermions are spinless, the pairing wave function is expected 
to have odd parity due to the antisymmetry requirement of the fermionic 
many body wavefunctions, and $p$-wave symmetry with the lowest angular 
momentum seems to be the leading candidate.
Nevertheless, we find that the configuration of $\Delta_A=-\Delta_B=\Delta$
exhibits the $f$-wave symmetry, {\it i.e.}, the rotation of $R_{\frac{\pi}{3}}$
is equivalent to flipping the sign of $\Delta$ as depicted in
Fig. \ref{fig:fwave} A.
In particular, the diagonal terms satisfy
\bea
R_{\frac{\pi}{3}}\Delta_{nn}(\vec k) R^{-1}_{\frac{\pi}{3}}
=-\Delta_{nn} (\vec k)
\eea
as exhibit in $F(\vec k)$ 
with three nodal lines of $k_x=0, k_y=
\pm k_x/\sqrt{3}$.
These are the same middle lines marked in Fig. \ref{fig:eigenvectors} 
along which the time reversal partners $\psi_n(\pm\vec k)$ 
have the same polar orbital configurations.
The pairing amplitude vanishes along these lines because interaction only
exists between two orthogonal orbitals.
On the other hand, maximal pairings  occur between $\vec K$ and $\vec K^\prime$ 
with opposite signs whose orbital configurations are 
orthogonal to each other.
In order words, the property of the reflection symmetry of the $p_{x,y}$ 
orbitals leads to vanishing of the intra-band pairing interaction along three 
nodal lines.
As a result, the pairing state favors the $f$-wave over the $p$-wave 
symmetries in order to match the nodal lines of the pairing interactions.
It is emphasized that unlike other examples of unconventional pairing 
in condensed matter systems, this $f$-wave pairing structure mainly 
arises from the non-trivial orbital configuration of band structures 
but with conventional interactions.

%-----------------------------------------------------------------------
\begin{figure}
\centering
\includegraphics{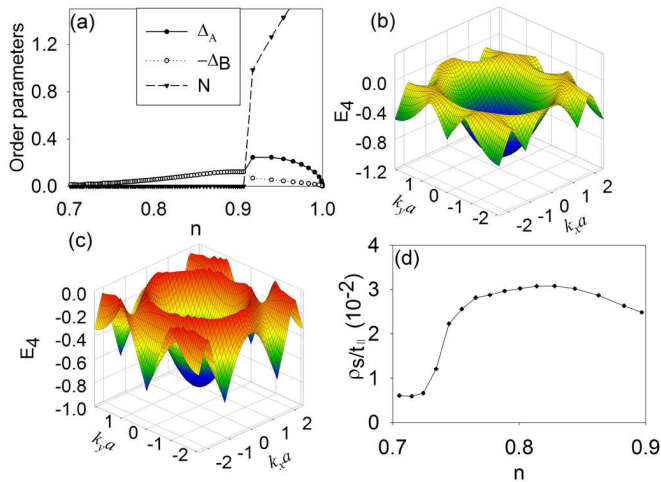}
\caption{\label{fig:sdgap} 
(Color online) Gap and superfluid density 
for the $f$-wave state at $U/t_\pp=2.2$ and $0.7<n<1.0$.
A) The pairing amplitude $\Delta_A=-\Delta_B=\Delta$ until $n\approx 0.9$. 
For $n>0.9$, the CDW order starts to coexist with the $f$-wave pairing superfluid, resulting  in $f$-wave supersolid state.; 
B) and C) the lowest energy branch of Bogoliubov excitations
at $n=0.89$  and $n=0.78$. 
For a better visual effect, only the negative eigenvalues are shown.  
D) The superfluid density $\rho_s$ v.s. $n$.
}
\label{fig:gap}
\end{figure}
%-----------------------------------------------------------------------

We next study the pairing strength $\Delta_A=-\Delta_B=\Delta$ 
in the weak coupling regime with $U/t_\pp=2.2$ 
in which the validity of the self-consistent mean-field theory is justified.
The lattice chiral symmetry ensures that these results are symmetric 
respect to the filling $n=1$ so that we only present the results for 
$n\leq 1$.
The case of $0<n<0.5$ corresponds to filling in the bottom flat band
in which each band eigenstate can be constructed as localized within
a single hexagon plaquette presented (see Ref. \cite{wu2007, wu2008}).
The attractive interaction drives all the plaquette states to touch
each other leading to phase separation until the flat band is 
filled, which will be discussed in a later publication.
In Fig. \ref{fig:gap} A, we plot $\Delta$ for $0.7<n<1.0$ which
corresponds to filling in the dispersive bands.
If can be seen that the pairing state remains pure $f$-wave until 
$n\approx 0.9$.
For $n>0.9$, the CDW order parameter $N$ becomes non-zero due to the 
strong nesting near half-filling, and it contributes to a $s$-wave component
to the inter-band pairing, resulting in $\Delta_A\neq -\Delta_B$ observed 
in Fig. \ref{fig:gap} A. 
Since the CDW order coexists with the pairing superfluid with dominant $f$-wave 
component, the phase in the region of $0.9<n<1.0$ is the $f$-wave supersolid 
state, which is also a novel phase not seen in other systems.

Due to the multi-band structure, this $f$-wave pairing state
remains fully gapped for general values of $U/t_\pp$ and $n$.
We plot the branch of the lowest energy Bogoliubov excitations for $U/t_\pp=2.2$
and the filling $0.7<n<1$ where the Fermi energy lies in band 2.
For $n$ close to 1 in Fig. \ref{fig:gap} B ($n=0.89$), 
the Fermi surfaces form two disconnected 
pockets around the $\vec K$ and $\vec K^\prime$ away from the 
nodal lines of $\Delta_{22} (\vec k)$, thus the spectra are gapped.
As $n$ is lowered in Fig. \ref{fig:gap} C ($n=0.78$), the Fermi surface 
becomes connected and intersects with the nodal lines of $\Delta_{22}$.
At these intersections, the system in general remains gapped because 
of the non-zero inter-band pairing $\Delta_{12}(\vec k)$.

The superfluid density $\rho_s$ is plotted in Fig. \ref{fig:gap} D
for $0.7<n<1.0$ and $U/t_\pp=2.2$, which is defined as
\bea
\rho_s \equiv \frac{\hbar^2}{2m^*} n_s = 
\lim_{\delta\theta\rightarrow 0}\frac{1}{2}\frac{\partial^2 E_{MF}}{\partial \delta \theta^2}
\eea
where $\delta\theta$ is the phase twist across the system boundaries.
The behavior of $\rho_s$ is very different from that of the pairing gap,
which is mostly determined by the states in the dispersive band.
$\rho_s$ reaches the maximal value of about $0.03t_\pp$ around $n=0.82$ because
the filling is close to the van Hove singularity
of the density of states, and then drops 
as we approach the Dirac points at $n=1$.
In two dimensional systems, the superfluidity develops below the 
Kosterlitz-Thouless (K-T) transition temperature 
$T_{KT}\approx \frac{\pi}{2} \rho_s$.

%---------------------------------------------------------------
\begin{figure}
\centering\epsfig{file=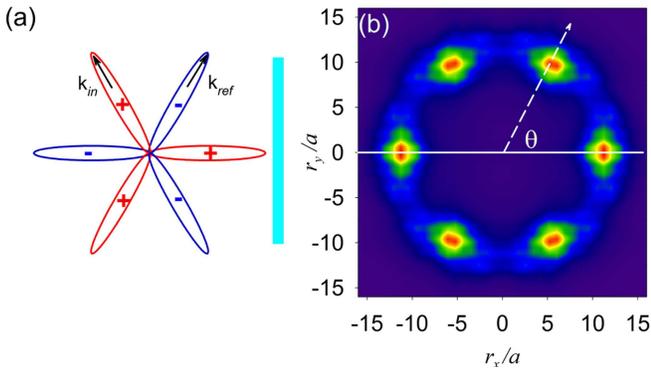,clip=1,width=\linewidth,angle=0}
\caption{\label{fig:andreev}
(Color online) (a)The zero energy Andreev bound states
appear on the boundary perpendicular to the anti-nodal directions.
(b) The zero energy LDOS in real space with the maximal values at
$\theta=n\pi/3$ is a signature of the $f$-wave pairing symmetry.}
\end{figure}
%---------------------------------------------------------------

The above mechanism for the $f$-wave pairing works in the weak
coupling systems where the non-trivial orbital band structure 
in momentum space is essential.
On the other hand, in the strong coupling limit, {\it i.e.}, 
$U$ is much larger than the band width, the Cooper pairing occurs 
in real space.
Two fermions on the same site form a Cooper pair which can tunnel
to neighboring sites. 
As explained in Ref. [\onlinecite{hung2009}], due to the odd parity of 
the $p$-orbitals, the Josephson coupling amplitude is positive,
which favors a phase difference of $\pi$ between two neighbors.
Furthermore, its amplitude is much smaller
than the CDW coupling because the $\pi$-bonding $t_\perp$ 
is much smaller than $t_\pp$.
As a result, the superfluid density is very small in the strong coupling
limit although the pairing gap is in the order of $U$.

%%%%%%%%%%%%%%%%%%%%%%%%%%%%%%%%%%%%%%%%%%%%%%%%%%%%%%%%%%%%%%%%%%%%%%%
\section{Zero energy Andreev Bound States}
\label{sect:andreev}

One of the most convincing proofs of the unconventional pairing is the 
existence of the zero energy Andreev bound states at boundaries
because of their phase sensitivity \cite{hu1994}.
Fig. \ref{fig:andreev} A depicts the situation of the boundary 
perpendicular to the antinodal lines of the intra-band pairing.
The scattering of the Bogoliubov quasi-particles changes 
momentum from ${\vec k}_{in}$ to $\vec {k}_{ref}$ along which
the pairing parameters switch the sign as
$\Delta_{nn}(\vec{k}_{in})=-\Delta_{nn}(\vec{k}_{ref})$,
which gives rise to zero energy Andreev bound states.
On the other hand, if the boundary is perpendicular to the nodal
lines, no phase changes occur and thus Andreev bound states vanish.

Naturally in the experimental systems with an overall confining 
trap, the circular edge boundary samples all the orientations.
We performed a real space Bogoliubov-de-Gennes calculation \cite{degennes} as
\bea
\left(\begin{array}{cc}
\hat{H}& \hat{\Delta} \\
\hat{\Delta}^* &-\hat{H}\\
\end{array}\right)\left(\begin{array}{c}
u_n(i)\\
v_n(i)\\
\end{array}\right)=E_n
\left(\begin{array}{c}
u_n(i)\\
v_n(i)\\
\end{array}\right),
\eea
where $i$ is the site index in the real space, $H=H_0+V_{ex}$, $V_{ex}=\frac{m}{2}\Omega^2 r_i^2$ is the harmonic confining
potential with parameters $m\Omega^2 a^2=0.001 t_\parallel$, and $r_i$ is the distance from the center of the confining potential to the site $i$.
In order to simulate better the realistic experimental situation, we also add the $\pi$-bonding term\cite{wu2007a}, the hopping with direction 
{\it perpendicular} to the bond direction, into $H_0$, and the hopping integral is estimated to be $t_{\perp}/t_\parallel=-0.1$.
The local density of states (LDOS) at energy $E$ is defined as
\bea
LD(\vec r,E)&=&\frac{1}{N}\sum_n \vert u_n(\vec r)\vert^2 \delta(E-E_n) \nn \\
&+& \vert v_n(\vec r)\vert^2 \delta(E+E_n),
\eea
where $N$ is the total number of lattice sites.
The numerical result of the zero energy $LD(\vec r,0)$ is depicted 
in real space in Fig. \ref{fig:andreev} B with the parameter values 
of $U/t_\pp=2.5$ and $\mu=-1.58$ which corresponds to the filling in
the center of the trap $n_c=0.88$.
Fig. \ref{fig:andreev} B shows that the $LD(\vec r,0)$ is non-zero
only at sites near the boundary, signaling the existence of the zero 
energy Andreev bound states with six-fold symmetry.
The LDOS maxima occur at angles $\theta=n\frac{\pi}{3} ~(n=1\sim 6)$  
which are perpendicular
to the antinodal directions shown in Fig. \ref{fig:andreev} A,
and the minima are located at $\theta =(n+\frac{1}{2})\frac{\pi}{3}$. 
As $U$ increases, the CDW order starts to coexist with the $f$-wave pairing, and
it introduces $s$-wave inter-band pairings into the system.
Consequently, the scattering process of Bogoliubov quasiparticles
described in Fig. \ref{fig:andreev}(a) no longer has a complete sign 
change in the pairing parameters as discussed above.
In this case, we still find the Andreev bound states with six-fold 
symmetry up to $U/t_\parallel=3$,
except having finite energy below the excitation gap instead of zero energy.

The zero energy Andreev bound states of this $f$-wave pairing systems have 
the similar topological origin as those in the superconductors
of the $p_x+i p_y$ symmetry.
However, there are important differences. 
The $f$-wave pairing symmetry presented here is real and anisotropic, 
which maintains time-reversal symmetry.
The edge responses are also anisotropic: the spectra of the zero energy 
Andreev state are strongest along the boundaries perpendicular to the 
anti-nodal direction.
On the other hand, the $p_x+ip_y$ pairing symmetry is complex. 
A spatial rotation operation on the $p_x+ip_y$ state is equivalent
to the change of a global pairing phase, thus its responses to 
all the boundaries with different directions are all the same.
Edge states appear regardless the directions of boundaries.
Furthermore, due to its breaking  time-reversal symmetry,
the edge states in the $p_x+ip_y$ states carry currents.

%%%%%%%%%%%%%%%%%%%%%%%%%%%%%%%%%%%%%%%%%%%%%%%%%%%%%%%%%%%%%%%%%%%%%%%
\section{Experimental realization and detection}
\label{sect:detect}

The experimental realization of this novel $f$-wave state is feasible.
To enhance the attractive interaction between spinless fermions,
we propose to use atoms with large magnetic moments, such as $^{167}$Er 
with $m=7\mu_B$ on which laser cooling has been performed 
\cite{mcclelland2006}.
Compared to another possibility to use the $p$-wave Feshbach resonance,
this method has the advantage to maintain the system stability.
The interaction between two fermions in $p_{x,y}$-orbitals reads
\bea
-U&=&\int d^3 \vec r_1 d^3 \vec r_2 V(\vec r_1-\vec r_2)
\Big \{ [\psi_{p_x}(\vec r_1) \psi_{p_y}(\vec r_2)]^2\nn \\
&-&\psi^*_{p_x}(\vec r_1) \psi^*_{p_y} (\vec r_2) \psi_{p_x} (\vec r_2)
\psi_{p_y} (\vec r_1) \Big \},
\eea
where $\psi_{p_x,p_y}$ are Wannier wavefunctions.
The overall dipole interaction can be made attractive by 
polarizing the magnetic moments parallel to the hexagonal plane,
which gives rise to 
$V(\vec r_1 -\vec r_2)=m^2 (1-3 \cos^2 \theta)/r^3$
where $r=|\vec r_1 -\vec r_2|$ and $\theta$ is the angle
between $\vec m$ and $\vec r_1-\vec r_2$.
We take the laser wavelength $\lambda\approx 600nm$, and then
the recoil energy $E_R=157 nK$.
As shown in the Supplementary Information, by choosing
$V_0/E_R=30$, the estimation shows that $U \approx 96 nK$
and $t_\pp \approx 0.2 E_R = 31nK$, and thus $U/t_{\pp}\approx 3$
as chosen above.
As shown in Fig. \ref{fig:gap} D, the maximal $T_{KT}\approx 
\frac{\pi}{2}(0.02 t_\pp) \approx 1\sim 2 nK$, which is within 
the experimentally accessible regime.

A successful detection of the zero energy Andreev bound states 
will be a convincing proof to the $f$-wave pairing state.
The radio-frequency ({\it rf}) spectroscopy has been an established 
tool to determine the pairing gap in cold 
atom systems \cite{chin2004,he2005}.
It also has a good spatial resolution \cite{schirotzek2008}, 
which makes the direct imaging of the spatial distribution 
of the zero energy Andreev bound states feasible.
The unique symmetry pattern of the zero energy Andreev
bound states localized at the trap boundary can be revealed
by identifying the locations of the zero energy spectrum
determined from the spatially resolved {\it rf} spectroscopy.
%Wei-Cheng: New senteces added
Even for the case in which the Andreev bound states are not at zero 
energy due to the CDW order, since they are the lowest energy states 
and below the excitation gap, one can still find that 
the lowest energy spectra appear near the six locations 
shown in Fig. \ref{fig:andreev}(b).

%%%%%%%%%%%%%%%%%%%%%%%%%%%%%%%%%%%%%%%%%%%%%%%%%%%%%%%%%%%%%%%%%%%%
\section{Conclusion and discussion}
\label{sect:conclusion}

In summary, we have proposed the realization of a novel $f$-wave 
pairing state with spinless fermions which has not been identified 
in solid state and cold atom systems before.
The key reason is the non-trivial $p$-orbital band structure of the
honeycomb lattice rather than the strong correlation physics, which
renders the above analysis controllable.
The $T_{KT}$ is estimated to reach the order of $1nK$ within experimental
accessibility.
The {\it rf} spectroscopy detection of the six-fold symmetry pattern
of the zero energy Andreev bound states along the circular boundary 
will provide a phase sensitive test of the $f$-wave symmetry.

\begin{acknowledgements}
C. W. acknowledges the Aspen Center for Physics where this work was
initiated. 
C. W. and W. C. L. are supported by NSF-DMR 0804775 and ARO-W911NF0810291.
S. D. S. is supported by ARO-DARPA and NSF-PFC.
\end{acknowledgements}

%\end{acknowledement}

%\setcounter{figure}{0} \setcounter{equation}{0}
%\renewcommand{\thefigure}{S\arabic{figure}}
%\renewcommand{\theequation}{S\arabic{equation}}

%\makeatletter \renewcommand\@biblabel[1]{S#1.} \makeatother
%%\renewcommand\citeform[1]{S#1} % parenthesized numbers [S1-S5]
%
%\renewcommand\refname{Supplementary References}

%%%%%%%%%%%%%%%%%%%%%%%%%%%%%%%%%%%%%%%%%%%%%%%%%%%%%%%%%%%%%%%
%%%%%%%%%%%%%%%%%%%%%%%%%%%%%%%%%%%%%%%%%%%%%%%%%%%%%%%%%%%%%%%%

%\newpage
\begin{appendix}
%\leftline{\Sup}

\section{Eigenvectors of the band Hamiltonian and the pairing matrix}
\label{sect:appendixA}

In this section, we present the spectra of the band Hamiltonian 
Eq. \ref{eq:ham0} and the pairing matrix $\Delta_{mn}$.
With the four-component spinor operator defined as
\bea
\hat \phi(\vec k)=(p_{Ax}(\vec k),p_{Ay}(\vec k), 
p_{Bx}(\vec k),p_{By}(\vec k))^T,
\eea
Eq. \ref{eq:ham0} becomes
\bea
H_0&=&t_\pp\sum_k \hat \phi^\dagger_m(\vec k)
H_{0,mn}(\vec k)  \hat \phi_n(\vec k),
\eea
where the matrix kernel $H_{0,mn}(\vec k)$ takes the structure as
{\small
\bea
\left(
\begin{array}{cccc}
0&0& \frac{3}{4} (e^{i \vec k \cdot \vec e_1} +e^{i \vec k \cdot \vec e_2})
&\frac{\sqrt 3}{4} (e^{i \vec k \cdot \vec e_1} -e^{i \vec k \cdot \vec e_2})
\\
0&0& \frac{\sqrt 3}{4} (e^{i \vec k \cdot \vec e_1} -e^{i \vec k
\cdot \vec e_2})&
\frac{1}{4} (e^{i \vec k \cdot \vec e_1} +e^{i \vec k \cdot \vec e_2})
+e^{i \vec k \cdot \vec e_3}\\
h.c.& & 0&0 \\
   & & 0&0
\end{array}
\right). \nn 
\eea
}

For each momentum $\vec k$, $H_0(\vec k)$ is diagonalized as 
\bea
H_{0,lm}(\vec k) U_{mn} = E_l U_{ln}.
\eea
The band eigen-operators are expressed as
\bea
\hat \psi_m (\vec k) = \hat \phi_n (\vec k) U_{nm} (\vec k).
\eea
The unitary matrix $U(\vec k)$
reads
\bea
U(\vec{k})=
\left(\begin{array}{c c c c}
a^*(\vec{k}) & e^{-i\frac{\theta_k}{2}}b(\vec{k}) & 
e^{-i\frac{\theta_k}{2}}b(\vec{k}) & a^*(\vec{k})\\
-b^*(\vec{k}) & e^{-i\frac{\theta_k}{2}}a(\vec{k}) 
& e^{-i\frac{\theta_k}{2}}a(\vec{k}) & -b^*(\vec{k})\\
a(\vec{k}) & e^{i\frac{\theta_k}{2}}b^*(\vec{k}) 
& -e^{i\frac{\theta_k}{2}}b^*(\vec{k}) & -a(\vec{k})\\
-b(\vec{k}) & e^{i\frac{\theta_k}{2}}a^*(\vec{k}) 
& -e^{i\frac{\theta_k}{2}}a^*(\vec{k}) & b(\vec{k})
\end{array}\right),  \nn \\ 
\eea
where
\bea
a(\vec{k})&=&\frac{f_{23}(\vec{k})-f_{31}(\vec{k})}{\sqrt{3N_0(\vec{k})}};
\ \ \
b(\vec{k})=\frac{f_{12}(\vec{k})}{\sqrt{N_0(\vec{k}})}; \nn \\
\theta_k&=&{\rm arg}\left(\sum_i e^{i\vec{k}\cdot\hat{e}_i}\right) \in [-\pi,\pi);
\nn \\
f_{ij}(\vec{k})&=& e^{i\vec{k}\cdot\hat{e}_i}-e^{i\vec{k}\cdot\hat {e}_j}; \nn \\
N_0(\vec{k})&=&\frac{8}{3} \Big[3-\sum_{1\le i<j\le 3} 
\cos\vec{k}\cdot(\hat{e}_i-\hat{e}_j )\Big].
\eea

Using the eigenvector matrix $U(\vec{k})$, the pairing  matrix
$\Delta_{mn}$ can be spelled out straightforwardly as:
\bea
\Delta_{mn}(\vec{k}) &=& \Delta_A\left[\hat{U}_{2m}(-\vec{k})\hat{U}_{1n}(\vec{k}) - \hat{U}_{1m}(-\vec{k})\hat{U}_{2n}(\vec{k})\right] \nn\\
                     &+& \Delta_B\left[\hat{U}_{4m}(-\vec{k})\hat{U}_{3n}(\vec{k}) - \hat{U}_{3m}(-\vec{k})\hat{U}_{4n}(\vec{k})\right]. \nn\\
\label{44gap}
\eea
For the solution of $\Delta_A=-\Delta_B=\Delta$, we have the pairing
matrix as
\bea
\Delta(\vec{k})=
\left(\begin{array}{c c c c}
\Delta_1(\vec{k}) & \Delta_2(\vec{k}) & \Delta_3(\vec{k}) & 0 \\
\Delta_2(\vec{k}) & -\Delta_1(\vec{k}) & 0 & -\Delta_3(\vec{k}) \\
-\Delta_3(\vec{k}) & 0 & -\Delta_1(\vec{k}) & \Delta_2(\vec{k}) \\
0&\Delta_3(\vec{k}) & \Delta_2(\vec{k}) & \Delta_1(\vec{k}) \\
\end{array}\right),
\eea
where the matrix elements read
\bea
\Delta_1(\vec{k})&=&i\frac{16\Delta}{\sqrt{3} N_0(\vec{k})}
\sin\frac{\sqrt{3}}{2}
k_x\left[\cos\frac{\sqrt{3}}{2}k_x - \cos\frac{3}{2}k_y\right],\nonumber\\[2mm]
\Delta_2(\vec{k})&=&i\frac{16\Delta}{3N_0(\vec{k})}\left[K_1(\vec{k})\sin\frac{\theta_k}{2}+K_2(\vec{k})\cos\frac{\theta_k}{2}\right],\nonumber\\[2mm]
\Delta_3(\vec{k})&=&\frac{16\Delta}{3N_0(\vec{k})}\left[K_1(\vec{k})\cos\frac{\theta_k}{2}-K_2(\vec{k})\sin\frac{\theta_k}{2}\right],\nonumber\\[2mm]
\eea
and
\bea
K_1(\vec{k})&=&4\cos^4\frac{k_y}{2} + (4\cos^2\frac{\sqrt{3}}{2}k_x-7)\cos^2\frac{k_y}{2}\nn \\
&-& \cos\frac{k_y}{2}\cos\frac{\sqrt{3}}{2}k_x + 
2 \sin^2\frac{\sqrt{3}}{2}k_x, \nn \\
K_2(\vec{k})&=&\sin\frac{k_y}{2}(\cos\frac{\sqrt{3}}{2}k_x-\cos\frac{k_y}{2})
\nn \\
&\times&[(2\cos\frac{k_y}{2}+\cos\frac{\sqrt{3}}{2}k_x)^2
+\sin^2\frac{\sqrt{3}}{2}k_x]. \ \ \
\eea

Both the band Hamiltonian Eq. \ref{eq:ham0} and the interaction
Eq. \ref{eq:int}
are invariant under the rotation $R_{\frac{\pi}{3}}$.
Applying $R_{\frac{\pi}{3}}$ to the band Hamiltonian matrix in momentum 
space $H_0(\vec k)$, it transforms as
\bea
H_0[\vec{k}^\prime=R_{\frac{\pi}{3}}(\vec k)]= R_{\frac{\pi}{3}}H_0(\vec{k})R^{-1}_{\frac{\pi}{3}}
\eea
where
$R_{\frac{\pi}{3}}$ is defined as
\be
R_{\frac{\pi}{3}}=
\left(
\begin{array}{cc}
0&r_{\frac{\pi}{3}}\\
r_{\frac{\pi}{3}}&0\\
\end{array}
\right),
\ee
and
\be
r_{\frac{\pi}{3}}=
\left(
\begin{array}{cc}
\cos\frac{\pi}{3}&-\sin\frac{\pi}{3}\\
\sin\frac{\pi}{3}&\cos\frac{\pi}{3}\\
\end{array}
\right).
\ee
Correspondingly, it can be shown directly that 
\bea
R_{\frac{\pi}{3}} \hat \psi_m (\vec k) 
=\sgn(m) \hat \psi_m(\vec k^\prime)
\eea
 with $\sgn(m)=-1$ for $m=1,2$ and $+1$ for $m=3,4$, 
and $R_{\frac{\pi}{3}} \Delta_{mn} (\vec k) R_{\frac{\pi}{3}}^{-1} 
=\sgn(n)\sgn(m) \Delta_{mn} (\vec k^\prime)$.

\section{Calculation of the on-site Hubbard interaction with magnetic
dipolar interaction}
\label{sect:appendixB}

\begin{figure}
\centering\epsfig{file=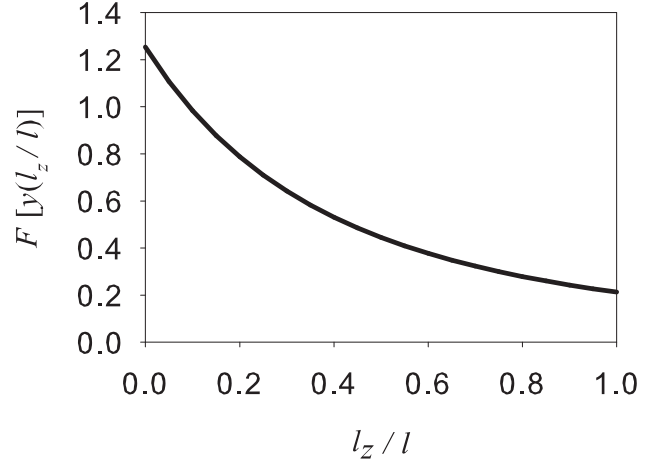, clip=1, width=\linewidth, angle=0}
\caption{$F[y(l_z/l)]$ as a function of $l_z/l$.
}
\label{fig:estiu}
\end{figure}

The optical potential around the center of each optical site can be
approximated as
\bea
V(\vec{r})=\frac{m}{2}\omega^2(r_x^2+r_y^2) + \frac{m}{2}\omega_z^2 r_z^2,
\eea
where we assume that the confinement in the $z$-axis is stronger 
so that $\omega_z>\omega$.
The wavefunctions for the $p_x$ and $p_y$ orbitals in this harmonic
potential are
\bea
\psi_{p_x}(\vec{r})=C^{\frac{1}{2}} r_x 
\exp\big\{-\frac{r_x^2+r_y^2}{2l^2}-\frac{r_z^2}{2l_z^2}\big \},\nonumber\\
\psi_{p_y}(\vec{r})=C^{\frac{1}{2}} r_y 
\exp\big\{-\frac{r_x^2+r_y^2}{2l^2}-\frac{r_z^2}{2l_z^2}\big\},
\eea
where $C=2l^{-4} l^{-1}_z \pi^{-3/2}, l=\sqrt{\hbar/m\omega}$, 
$l_z=\sqrt{\hbar/m\omega_z}$, and $l_z < l$.

With these wavefunctions, the on-site interaction can be evaluated
with the direct and exchange terms:
\bea
-U&=&\int d^3 \vec r_1 d^3 \vec r_2 V(\vec r_1-\vec r_2)
\Big \{ [\psi_{p_x}(\vec r_1) \psi_{p_y}(\vec r_2)]^2\nonumber\\[2mm]
& &-\psi^*_{p_x}(\vec r_1) \psi^*_{p_y} (\vec r_2) \psi_{p_x} (\vec r_2)
\psi_{p_y} (\vec r_1)\Big \}\nonumber\\[2mm]
&=&C^2\int d^3 \vec{r} d^3 \vec{R}~ V(\vec{r}) 
\left(\frac{R_x^2\,r_y^2 + R_y^2\,r_x^2}{2}\right) \nn \\
&\times& F_1(\vec{r})F_2(\vec{R}),
\label{computeu}
\eea
where we have introduced the relative coordinate 
$\vec{r}=\vec{r}_1-\vec{r}_2$ and the center of mass coordinate
$\vec{R}=(\vec{r}_1+\vec{r}_2)/2$, and defined
\bea
F_1(\vec{r}) &=&\exp\big\{-\frac{ r_x^2+r_y^2}{2l^2}
-\frac{r_z^2}{2l_z^2}\big\}, \nn \\
F_2(\vec{R})&=& \exp\big\{-\frac{2(R_x^2+R_y^2)}{l^2}
-\frac{2R_z^2}{l_z^2}\big\}.
\eea

We propose to use fermionic atoms with large magnetic dipole moments
$\vec m$.
By polarizing the magnetic moments with an external magnetic field,
the anisotropic interaction reads
\bea
V(\vec{r})=\frac{\mu_0}{4\pi}\frac{m^2 [1-3(\hat{m}\cdot\hat{r})]^2} {r^3}
\eea
where $\hat{m}= \vec{m}/m$ and $\hat{r}= \vec{r}/r$, respectively.
We assume the polarization angle between $\vec{m}$ and the $xy$-plane
is $\theta$, and perform the Gaussian integrals in Eq. \ref{computeu}.
The result is expressed analytically as
\bea
-U&=&\frac{\mu_0}{4\pi}\frac{m^2}{2l^3}
(1-\frac{3}{2}\cos^2\theta)F(y),
\eea
where 
\bea
F(y)&=&\sqrt{\frac{2}{\pi}}(1+y)^{3/2} [(3y+1)
\tan^{-1}(\frac{1}{\sqrt{y}}) \nn \\
&-&3\sqrt{y} ],
\eea
and $y$ is a function of $l_z/l$ as 
$y(l_z/l)= l_z^2/(l^2-l_z^2)$.
$F[y(l_z/l)]$ is a monotonic decreasing function of $l_z$ as shown 
in Fig. \ref{fig:estiu}, which reflects that the wavefunctions 
with a larger $l_z$ have smaller overlaps with each other.

The largest attractive interaction can be 
achieved by polarizing the magnetic moments in the $xy$-plane.
If we use the fermionic atom of $^{167}$Er with $m=7\mu_B$ and 
the laser beam with the wavelength $\lambda\approx 600nm$ 
to construct the honeycomb optical lattice, 
the recoil energy $E_R=h^2/2m\lambda^2=157 nK$ 
and $l\approx 93 (\frac{E_R}{V_0})^{1/4}$nm. 
Compared to the realistic band structure calculation with $V_0/E_R=30$
in Ref. \cite{wu2008}, $t_\pp$ is fitted as 
$t_\parallel\approx 0.2 E_R=31$nK and $l\approx 40$nm.
With the choice of $l_z/l=0.2$, $U$ reaches reach $96$nK, 
thus $U/t_\parallel\approx 3$ employed in this paper is achieved.
Recently, a large number of the fermionic atoms $^{163}$Dy with $m=10\mu_B$ has been successfully 
cooled and trapped\cite{lu2010}. 
For this atom, the recoil energy $E_R=h^2/2m\lambda^2=161 nK$, 
and $U/t_\parallel\approx 3$ can be achieved with the choice of $l_z/l=0.55$.

\end{appendix}

%\bibliographystyle{prsty}
%\bibliography{orbital,spin32,topo}

\end{document}